\documentclass[aps,twocolumn,showpacs,amsmath,amssymb,superscriptaddress,final,prb,10pt]{revtex4-1}
\usepackage[latin1]{inputenc}   
\usepackage{dcolumn}  
\usepackage{nicefrac}   
\usepackage[dvips]{color} 

\usepackage{charter}

\definecolor{red}{rgb}{0.85,.1,0}
\definecolor{cblue}{named}{CadetBlue}
\usepackage{graphicx}          
\DeclareGraphicsExtensions{.eps,.png,.pdf}
\newcommand{\LOFBS}{La\-O$_{0.5}$F$_{0.5}$\-Bi\-S$_{2}$}
\newcommand{\NOFBS}{Nd\-O$_{1-x}$F$_{x}$\-Bi\-S$_{2}$}
\newcommand{\musr}{$\mu$SR}
\newcommand{\TF}{TF-$\mu$SR}
\newcommand{\ZF}{ZF-$\mu$SR}
\begin{document}
\title{\textit{s}-wave pairing in the optimally-doped LaO$_{0.5}$F$_{0.5}$BiS$_{2}$ superconductor}
\author{G.~Lamura}
\affiliation{CNR-SPIN and Universit\`a di Genova, via Dodecaneso 33, I-16146 Genova, Italy}
\author{T.~Shiroka}
\email[Corresponding author: ]{tshiroka@phys.ethz.ch}
\affiliation{Laboratorium f\"ur Festk\"orperphysik, ETH-H\"onggerberg, CH-8093 Z\"urich, Switzerland}
\affiliation{Paul Scherrer Institut, CH-5232 Villigen PSI, Switzerland}

\author{P.~Bonf\`a}
\affiliation{Dipartimento di Fisica e Scienze della Terra and Unit\`a CNISM di Parma, I-43124 Parma, Italy}
\author{S.~Sanna}
\affiliation{Dipartimento di Fisica and Unit\`a CNISM di Pavia, I-27100 Pavia, Italy}
\author{R.~{De~Renzi}}
\affiliation{Dipartimento di Fisica e Scienze della Terra and Unit\`a CNISM di Parma, I-43124 Parma, Italy}
\author{C.~Baines}
\affiliation{Paul Scherrer Institut, CH-5232 Villigen PSI, Switzerland}
\author{H.~Luetkens}
\affiliation{Paul Scherrer Institut, CH-5232 Villigen PSI, Switzerland}
\author{J.~Kajitani}
\affiliation{Department of Electrical and Electronic Engineering, Tokyo Metropolitan University, Hachioji, Tokyo 192-0397, Japan}
\author{Y.~Mizuguchi}
\affiliation{Department of Electrical and Electronic Engineering, Tokyo Metropolitan University, Hachioji, Tokyo 192-0397, Japan}
\affiliation{National Institute for Materials Science, Tsukuba, Ibaraki 305-0047, Japan}
\author{O.~Miura}
\affiliation{Department of Electrical and Electronic Engineering, Tokyo Metropolitan University, Hachioji, Tokyo 192-0397, Japan}
\author{K.~Deguchi}
\affiliation{National Institute for Materials Science, Tsukuba, Ibaraki 305-0047, Japan}
\author{S.~Demura}
\affiliation{National Institute for Materials Science, Tsukuba, Ibaraki 305-0047, Japan}
\author{Y.~Takano}
\affiliation{National Institute for Materials Science, Tsukuba, Ibaraki 305-0047, Japan}
\author{M.~Putti}
\affiliation{CNR-SPIN and Universit\`a di Genova, via Dodecaneso 33, I-16146 Genova, Italy}
\date{\today}
\begin{abstract}
We report on the magnetic and superconducting properties of La\-O$_{0.5}$F$_{0.5}$\-Bi\-S$_{2}$ by means of zero- (ZF) and transverse-field (TF) muon-spin spectroscopy measurements ($\mu$SR). Contrary to previous results on iron-based superconductors, measurements in zero field demonstrate the absence of magnetically ordered phases. TF-$\mu$SR data give access to the superfluid density, which shows a marked 2D character with a dominant $s$-wave temperature behavior. The field dependence of the magnetic penetration depth confirms this finding and further suggests the presence of an anisotropic superconducting gap.
\end{abstract}
\pacs{76.75.+i,74.25.Ha, 74.20.Rp, 74.25.Op}
\maketitle
Following the discovery of bulk superconductivity in Bi$_4$O$_4$S$_3$,\cite{MizuBiOS2012,Awana2012,Shruti2013} \LOFBS,\cite{Mizuguchi2012} and \NOFBS \cite{MizuNd2012}, with superconducting critical temperatures ranging from 4 up to 10 K, BiS$_2$ layered compounds are increasingly 
attracting the interest of the scientific community. Their structure consists of a charge reservoir layer (LnO, where Ln is a lanthanoid) alternated with BiS$_2$ planes that play the same role as CuO$_2$ planes in cuprates or the FeAs and Fe(Se,Te) planes in iron-based superconductors (IBS). Differently from both cuprates and IBS, where the conduction bands at the Fermi energy are dominated by 3$d$ states, in BiS$_2$ systems the main contribution comes from the bismuth 6$p$ orbitals, which are spatially extended and hybridized with the 3$p$ orbitals of sulfur.\cite{Usui2012,Wan2013,Li2013} This hybridization increases the delocalization of the charge carriers, but it is rather anisotropic.
Hence, Bi- and S-derived conduction bands show almost no dispersion along the $k_z$ direction,\cite{Usui2012,Wan2013} hinting at a marked 2D character also for the superconducting properties. 
Superconductivity in BiS$_2$ layered compounds occurs by electron doping the Bi 6$p$ bands. In particular, both tight-binding models\cite{Usui2012} and density functional calculations\cite{Wan2013} indicate that in \LOFBS\ four bands cross the Fermi level giving rise to a two-dimensional Fermi surface strongly nested at $(\pi,\pi,0)$ wave vector. However, there is a general lack of consensus regarding the consequences of nesting. On one hand, nesting is supposed to enhance the electron-phonon coupling, thus favoring conventional BCS superconductivity.\cite{Usui2012,Wan2013,Yildirim2013,Li2013} Indeed, a very recent penetration-depth measurements by a tunnel-diode technique on Bi$_4$O$_4$S$_3$ polycrystals indicates a conventional $s$-wave type superconductivity in the strong electron-phonon coupling limit.\cite{Shruti2013} On the other hand, as for IBS, a strongly nested Fermi surface could also imply that electronic correlations, rather than electron-phonon coupling, play a major role in determining the pairing mechanism.\cite{Zhou2013,Dagotto2013} In such a case, spin-fluctuation mediated pairing interactions could result either in an extended  $s$-wave or in a $d$-wave symmetry for the gap function.\cite{Zhou2013,Usui2012} Finally, because of the remarkable similarity with the  Sr$_2$RuO$_4$ band structure, a spin-triplet $p$-wave pairing has also been considered.\cite{Zhou2013,Usui2012}

Here we report on a study of the magnetic and superconducting properties of \LOFBS, mostly via muon-spin spectroscopy. 
\ZF\ shows that, contrary to iron-based superconductors,\cite{Johnston2010} there are no ordered magnetic phases in this 
material. On the other hand, \TF\ in the SC phase can track the evolution of the superfluid density with temperature, 
as well as that of the muon-spin depolarization rate with the applied field. Both these dependencies hint at a fully-gapped \textit{conventional} BCS picture of superconductivity in \LOFBS.

Eight polycrystalline samples were prepared as pellets\footnote{Each sample has a disc shape of 5 mm in diameter and 1.5 mm in thickness.} by using standard solid-state reaction techniques. Subsequent annealing took place at 600${}^{\circ}$C under a pressure of 2 GPa, realized by means of a cubic-anvil high-pressure experimental setup (see Ref.~\onlinecite{Mizuguchi2012}). All samples were firstly characterized by x-ray scattering and dc magnetometry, whose results (not shown here) confirm both the absence of impurity phases and the bulk nature of superconductivity. The critical temperature, $T_c=10$ K, is in agreement with data previously reported for samples having the same origin.\cite{Mizuguchi2012,DeguchiEPL,Kotegawa2012}
\musr\ measurements were performed by using both the General-Purpose Spectrometer (GPS) and the Low-Temperature Facility (LTF) on the $\pi$M3  beamline  of  the  Swiss  Muon  Source  at  the Paul  Scherrer  Institute,  Villigen,  Switzerland. Given the different beam cross 
sections, a single sample was sufficient at GPS, whereas a mosaic of the remaining seven samples was required for the very 
low-temperature measurements ($T < 1.7$ K) at LTF.\\
\begin{figure}[tbh]
\centering
\includegraphics[width=0.4\textwidth]{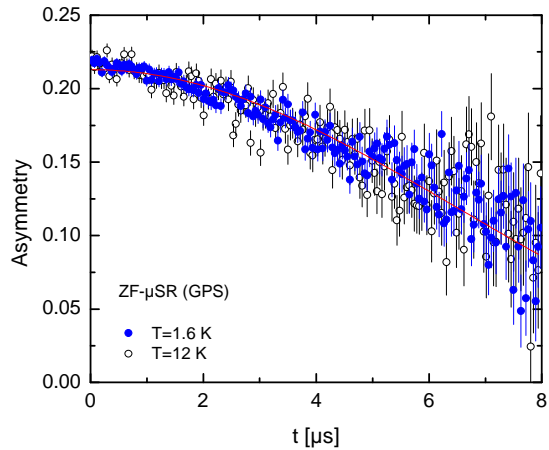}  
\caption{\label{fig:ZF_asym}(Color online) ZF-\musr\ asymmetry data collected above and below $T_c$, shown using a binning factor of 45. Continuous lines are best fits. See text for details.}
\end{figure}
By means of ZF-\musr\ we checked first for the presence of a possible hidden magnetic order in the BiS$_2$ planes. Magnetism coexisting with superconductivity is observed in underdoped pnictides,\cite{Sanna2009,Sanna2010,Johnston2010} but magnetic fields may also appear in superconductors with spin-triplet pairing, as observed in Sr$_2$RuO$_4$.\cite{Luke1998,Shiroka2012} In Fig.~\ref{fig:ZF_asym} the ZF time-dependent asymmetry below and above the superconducting transition is shown: both data sets were well fitted by a \textit{pure} Gaussian Kubo-Toyabe model, indicating the presence of small, broad static magnetic fields on the muon site due only to randomly-oriented nuclear magnetic dipole moments.\cite{Yaouanc2011} The Gaussian broadening $\Delta_{\mathrm{KT}}$ was found to be equal to 0.1145(1) and 0.1163(2) $\mu$s$^{-1}$ at 1.6 and 12 K, respectively. Thus, contrary to underdoped iron-based superconductors, no static magnetism of electronic origin could be detected in the Bi-based superconductor. At the same time, the small and $T$-independent broadening makes the presence of time-reversal symmetry broken superconductivity highly unlikely.\\
\TF\ measurements were performed by field cooling the samples in the mixed state. Taking into account the critical field values  $B_{c1}(\mathrm{2\,K}) = 1.3\pm0.2$ mT\footnote{See Supplemental Material (SM) for the measurements of the first critical field at 2 K and the T-dependent ZFC dc-susceptibility, both carried out on the same sample under test at the GPS facility.} and $B_{c2}(0)\simeq 10$ T,\cite{Mizuguchi2012} the magnetic field was set to 0.07 and 0.3 T on GPS (in the 1.7--12 K range) and to 0.07 T on the LTF spectrometer (in the 0.02--1.7 K range). In Fig.~\ref{fig:FFT}(a) we show the TF muon-spin precession recorded at 1.7 and 12 K in a 0.3-T magnetic field. The time-dependent asymmetry was fitted by using the equation:
\begin{align}
\label{eq:spin_prec}
A(t)&= A_0  \exp \left(-\frac{\sigma^{2}_{n}+\sigma^{2}_{sc}}{2} \,t^2\right) \cos( \gamma_{\mu} B_{\mu}  t + \phi),
\end{align}
where $A_0$ is the initial asymmetry, $\gamma_{\mu}=2\pi\times 135.53$ MHz/T is the muon gyromagnetic ratio, $B_{\mu}$ is the local field probed by the implanted muons, and $\phi$ is the initial phase. Here $\sigma_{n}$ represents the muon-spin relaxation rate induced by the nuclear dipolar moments, whereas $\sigma_{sc}$ is proportional to the second moment of the magnetic field distribution due to the superconducting vortex lattice. Above $T_c$, where $\sigma_{sc}=0$, we find that $\sigma_{n} =  0.119(2)$ $\mu$s$^{-1}$, a value which 
was kept fixed in all the subsequent fit iterations.\footnote{In the specific case, for the LTF measurements a further term $A_{bg}=A_{\mathrm{Ag}} \cos(\gamma_{\mu} B_{\mathrm{ext}} t + \phi) \exp \left(-\lambda_{\mathrm{Ag}} t\right)$ has to be added to take into account the background contribution from the Ag sample holder (Ag plate no.\ 249 with $\lambda_{\mathrm{Ag}} = 0.025$ $\mu$s$^{-1}$).} The chosen fitting function [Eq.~(\ref{eq:spin_prec})] derives from the assumption that the local field distribution $P(B)$ due to the vortex lattice is a single Gaussian line (typically observed in powders of anisotropic type-II superconductors), contrary to the asymmetric field distribution theoretically expected and observed in good single crystals.\cite{Sonier2000}
\begin{figure}[tbh]
\centering
\includegraphics[scale=0.65]{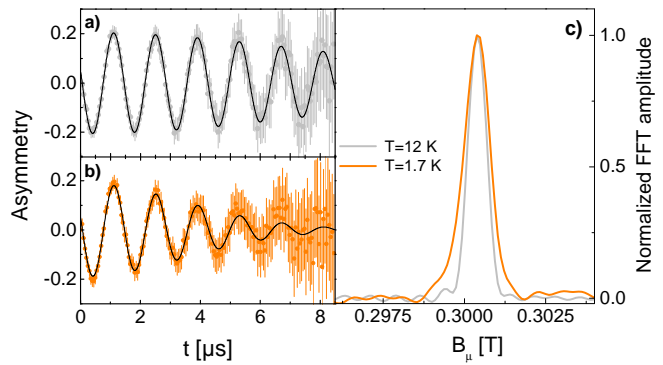}  
\caption{\label{fig:FFT}(Color online) TF-\musr\ asymmetry data collected above (a) and below (b) $T_c$ at an applied field of 0.3 T. For the sake of clarity the time-dependent asymmetry is represented in a 40-MHz rotating frame. (c) FFT real amplitude of the data shown in (a) and (b). Data processed using a Gaussian 7-$\mu$s time filter.}
\end{figure}
The choice of a single Gaussian is justified by the small additional relaxation value in the SC phase (indicative of a large penetration depth $\lambda$), comparable to that arising from the nuclear moments.\cite{Khasanov2008} As a result, the real part of the FFT signal [$\varpropto P(B)$], below and above $T_c$ [Fig.~\ref{fig:FFT}(c)], is always nearly symmetric. A large $\lambda$ implies also the occurrence of a tiny diamagnetic shift which can be appreciated only in Fig.~\ref{fig:shiftdia}, where the temperature dependence of the local field-shift $\delta B_{\mu}=B_{\mu}(T)-B_{\mu}(9.2K)$ together with the rescaled susceptibility (solid curve) are shown.
\begin{figure}[tbh]
\centering
\includegraphics[width=0.45\textwidth]{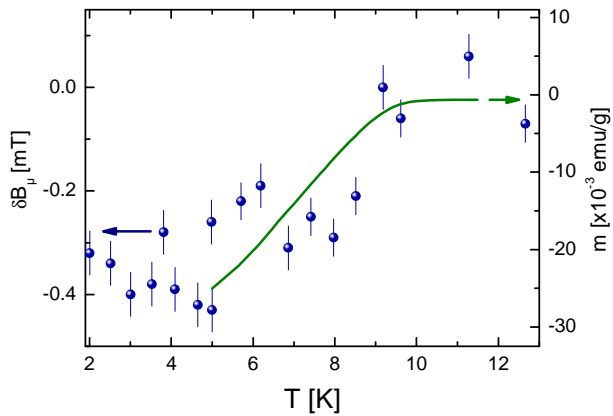}  
\caption{\label{fig:shiftdia}(Color online) Field-cooled (FC) dc magnetization (solid curve, right scale) and 
temperature dependence of the internal field shift $\delta B_\mu$ (solid circles, left scale), as from TF-\musr\ data.}
\end{figure}
The behavior of the square of the superconducting depolarization rate $\sigma_{sc}^2$, proportional to the second moment of the internal field distribution, is shown in Fig.~\ref{fig:sigmaB} as a function of the applied field.  Interestingly,  $\sigma_{sc}^2$ seems to increase linearly at low applied fields, but saturates above 0.2 T. This behavior is expected for hexagonal vortex lattices with a large Ginzburg-Landau parameter ($\kappa\ge 70$) at low fields $h = H/H_{c2}\ll 1$.\cite{Brandt03,Yaouanc2011}
\begin{figure}[tbh]
\centering
\includegraphics[width=0.4\textwidth]{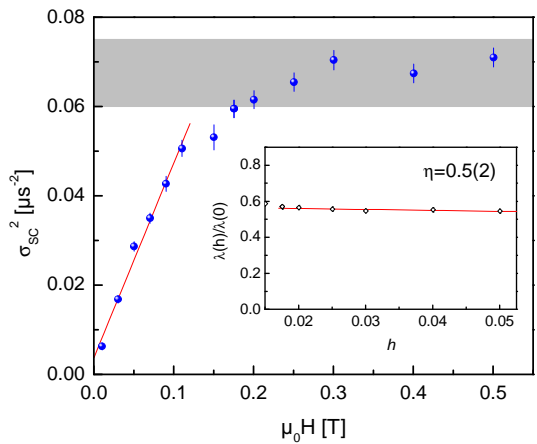}  
\caption{\label{fig:sigmaB}(Color online) Field dependence of the square of superconducting depolarization rate. The red line represents a linear fit in the low-field region. The grey area delimits the region of validity of Eq.~(\ref{eq:lambda}). Inset: field dependence of $\lambda(h)/\lambda(0)$ as extracted from the data in the grey region by means of Eq.~(\ref{eq:lambda}).}
\end{figure}
In this case the magnetic penetration depth $\lambda$ is related to $\sigma_{sc}$ by the equation:\cite{Brandt03} 
\begin{align}
\label{eq:lambda}
\sigma_{sc}=\gamma_{\mu} \, (0.00371)^{1/2} \frac{\phi_0}{\lambda_{\mathrm{eff}}^2}
\end{align}
where $\phi_0$ is the quantum of magnetic flux and $\lambda_{\mathrm{eff}}$ is the effective magnetic penetration depth. By using Eq.~(\ref{eq:lambda}), the estimated value for the latter at 0.3 T results $\lambda_{\mathrm{eff}}(\mathrm{1.7\,K})= 637\pm 4$ nm. Since \LOFBS\ is expected to be a highly anisotropic system, we can derive the in-plane (BiS$_2$) magnetic penetration depth by considering that in anisotropic polycrystalline samples $\lambda_{\mathrm{eff}}=3^{1/4} \lambda_{ab}$,\cite{Fesenko1991} which implies $\lambda_{ab}(\mathrm{1.7\,K}) = 484 \pm 3$ nm.

\begin{figure}[tbh]
\centering
\includegraphics[width=0.42\textwidth]{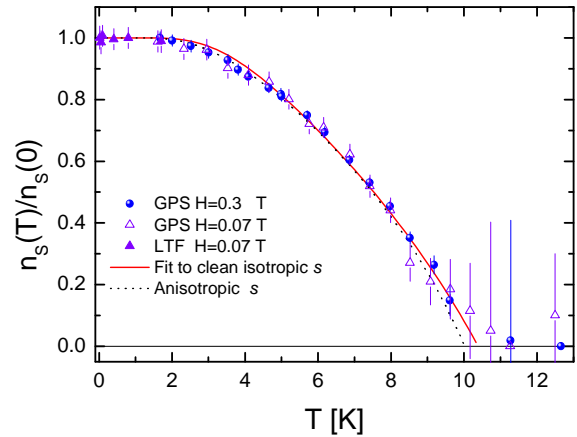}  
\caption{\label{fig:ns}(Color online) Temperature dependence of the normalized superfluid density measured at $\mu_0 H = 0.3$ T (full circles) and at 0.07 T (open triangles). Low-temperature ($T < 1.7$ K) LTF data at 0.07 T are shown by solid triangles. The full and the dotted black lines represent best fits to an $s$-wave gap in the clean limit and a numerical calculation with an $s$-wave anisotropic gap using the parameters reported in table~\ref{tab:fits}.} 
\end{figure}

By taking the zero-temperature limit of the coherence length extracted from the upper critical field measured in samples from the same origin,  $\xi(0)=[\phi_0/2 \pi B_{c2}(0)]^{1/2}\approx5.7$ nm,\cite{Mizuguchi2012} the resulting Ginzburg-Landau parameter is  $\kappa=\lambda_{ab}/\xi\approx85$. This value validates \textit{a posteriori} the choice of the above-mentioned theory [Eq.~(\ref{eq:lambda})] \cite{Brandt03} for extracting the magnetic penetration depth.
In the inset of Fig.~\ref{fig:sigmaB} the normalized effective penetration depth $\lambda(h)/\lambda(0)$ measured at 1.7 K is plotted as a function of the reduced magnetic field $h=B/B_{c2}$, limited to the experimental points corresponding to the grey region. It has been shown that in type-II superconductors $\lambda$ is sensitive to the presence of quasiparticle excitations due to gap nodes, double gaps or anisotropic gaps. In particular, the slope $\eta=|\mathrm{d}[\lambda(h)/\lambda(0)]/\mathrm{d}h|$ is mostly greater than 1 for a $d$-wave gap, it is around unity for double and anisotropic gaps, but it is almost zero for isotropic $s$-wave superconductors.\cite{Kadono2004} In our case, $\eta \simeq 0.5$ suggests the presence of an anisotropic $s$-wave gap or of a double $s$-wave gap.
Figure~\ref{fig:ns} shows the temperature dependence of the normalized supercarrier density $n_s(T)=\lambda_{ab}(T,H)^{-2}/\lambda_{ab}(\mathrm{1.7\,K},H)^{-2}=\sigma_{sc}(T)/\sigma_{sc}(0)$ calculated at 0.07 and 0.3 T: both data sets merge together thus confirming the fully gapped character of this superconductor.
It is relevant to note that: ($i$) $n_s(T)$ is \textit{temperature independent below 2.3 K}, thus ruling out gap nodes. ($ii$) The presence of a double gap is highly improbable too. At high fields the smaller gap would be suppressed before the larger one, 
giving rise to an (obviously missing) peak in Fig.~\ref{fig:sigmaB}, between the linear slope and the plateau. In addition, the low-$T$ contribution of the smaller gap to the superfluid density [Fig.~(\ref{fig:ns})] would have shown up as a kink, suppressed at higher applied fields,\cite{Khasanov2007} whereas both data sets follow closely a single-gap behavior. Consequently, we limit our discussion to the single $s$-wave gap picture. In this case the superfluid density in the clean local limit has the form\cite{TINK,Khasanov2008II}
\begin{align}
\label{eq:ns_clean}
n_s(T)=1+\frac{1}{\pi}\int_0^{2\pi}\!\!\!\int_{\Delta(T,\phi)}^{\infty} \left(\frac{\partial f}{\partial E} \right) \frac{E\,\mathrm{d}E\,\mathrm{d}\phi}{\sqrt{E^2-\Delta(T,\phi)^2}} 
\end{align}
where $f(E)$ is the Fermi function, $\phi$ is an angle spanning the supposed cylindrical Fermi surface and $\Delta(T,\phi)=\Delta_0 \delta(T) g(\phi)$ is the gap function. In particular, $\Delta_0$ is the maximum gap amplitude, $\delta(T)=\tanh\{1.82  [1.018(T_c/T-1)^{0.51}]\}$ is a widely accepted approximation for the temperature dependence of the gap;\cite{Khasanov2008II} $g(\phi)=1$ for an isotropic $s$-wave gap, or $g(\phi)=[1+a \,\cos(4 \phi)]/(1+a)$ for an anisotropic gap,\cite{Khasanov2007} 
with $\Delta_{\mathrm{min}}/\Delta_{\mathrm{max}}=(1-a)/(1+a)$ the ratio between the minimum and maximum gap amplitudes.
In Fig.~\ref{fig:ns} we report a fit to a single-gap model in the clean limit and a calculation for an anisotropic $s$-wave gap. For the sake of clarity, we do not show the fit to the single-gap model in the full dirty limit,\cite{TINK} the latter being indistinguishable from the clean limit. All the relevant fit and calculation parameters are summarized in Table~\ref{tab:fits}.
We remark that: ($i$) the full dirty limit is very unlikely because it gives a too small $2\Delta/k_\mathrm{B} T_c$ ratio. Assuming the predicted\cite{Wan2013} values for the electron-phonon coupling $\lambda=0.85$ and the logarithmic averaged phonon frequency $\omega_{ln}=22.4$ meV, the expected ratio is $2\Delta/k_\mathrm{B} T_c=3.7$.\cite{Carbotte1990} ($ii$) The fit to the $s$-wave model in the clean limit is closer to the predicted value since it gives $2\Delta/k_\mathrm{B} T_c=3.4$, that is compatible with a BCS model in the weak-coupling limit, contrary to the above mentioned theoretical predictions of a moderately strong coupling.\cite{Wan2013}  For this reason and considering that the parameter $\eta$ is not strictly zero, we cannot rule out also the possibility of an anisotropic $s$-wave gap. Indeed in Fig.~\ref{fig:ns} we show that the superfluid density calculated within an anisotropic $s$-wave gap model with an angular-averaged value $2\Delta/k_\mathrm{B} T_c=3.74$ is in very good agreement with the experimental data. We should recall that in extreme type-II superconductors with highly diluted superfluid density the presence of some anisotropy in the gap function is quite common.\cite{Brandow2003}\\
\begin{table}[tbh]
\caption{\label{tab:fits} Superconducting parameters as derived from the fits reported in Fig.~\ref{fig:ns}.}
\begin{ruledtabular}
\begin{tabular}{ccccc}
Gap model & $\Delta_0$ (meV) & $a$ & $2 \Delta_0/ k_\mathrm{B} T_c$ & $T_c$ (K)\\ \hline
$s$ clean      & 1.47 $\pm$0.03 &   -   & 3.4$\pm$0.2 & 10.4$\pm$0.1 \\
$s$ dirty      & 1.13 $\pm$0.02 &   -   & 2.6$\pm$0.1 & 10.5$\pm$0.1 \\
anisotropic $s$& 2.295          & 0.425 &   3.74\footnote{Value obtained by averaging the asymmetric gap over $[0, 2\pi]$.}    &      10     \\
\end{tabular}
\end{ruledtabular}
\end{table}
Finally, it is worth noticing that our $\lambda_{ab}(0)$ value is comparable with those found in alkali-metal organic-solvent intercalated 
iron-selenide superconductors.\cite{Biswas2013} By considering an interlayer distance equal to the $c$-axis parameter\cite{Mizuguchi2012} we can estimate a 2D superfluid density $d / \lambda^2_{ab}(0) \sim 0.0056$ $\mu$m$^{-1}$. This value places  \LOFBS\ close to the 
iron-chalcogenide superconductors in the  Uemura's plot of 2D superfluid density, as reported in Ref.~\onlinecite{Biswas2013}. 
This proximity is an important indication in favor of the 2D character of this compound and of its strong similarities with the cuprates 
and iron-chalcogenide superconductors.\\ \\
In conclusion, we have presented \musr\ results on the newly discovered \LOFBS\ superconductor. The temperature behavior of its superfluid density shows a marked $s$-wave character, with $2\Delta/k_\mathrm{B} T_c$ very near to the value expected for a phonon-mediated pairing,  with possibly an anisotropic gap. The high value of the Ginzburg-Landau parameter, $\kappa(0)\simeq85$, places this compound in the extreme type-II superconductor family. Finally, the in-plane magnetic penetration depth $\lambda_{ab}(\mathrm{1.7\,K})=484\pm3$ nm  
indicates a very dilute superfluid density, typical of systems with an almost 2D character.
\begin{acknowledgments}
We are grateful to A.\ Amato for the instrumental support and for the access to LTF for the low-temperature measurements. T.S.\ acknowledges support from the Schweizerische Nationalfonds zur F\"orderung der Wissenschaftlichen Forschung (SNF) and the NCCR research pool MaNEP of SNF.  G.L.\ and M.P.\ are grateful to M.R.\ Cimberle for fruitful discussions and acknowledge support from the FP7-EU project SUPER-IRON (No. 283204). S.S. acknowledges the financial support from Fondazione Cariplo (research grant no. 2011-0266).
\end{acknowledgments}
\section*{SUPPLEMENTAL MATERIAL}
We present here important additional magnetic characterization data regarding  the same sample measured via $\mu$SR at GPS facility.
\begin{figure}[tbh]
\centering
\includegraphics[width=0.42\textwidth]{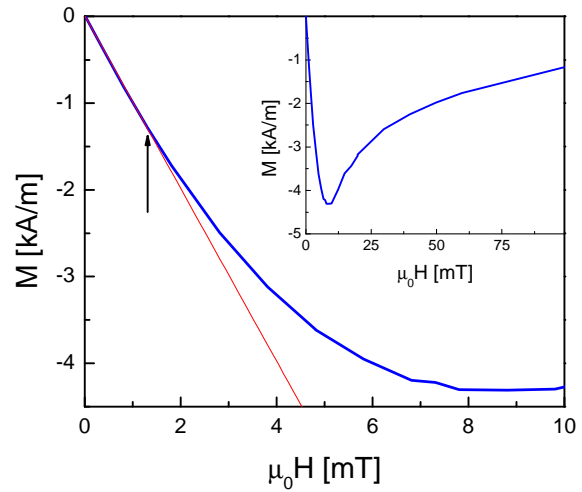}  
\caption{\label{fig:bc1}(Color online) Field dependence of magnetization taken at 2 K. The arrow indicates the field where $M(H)$ departs from linearity. The inset shows the same data over an extended field interval.}
\end{figure}
\begin{figure}[tbh]
\centering
\includegraphics[width=0.43\textwidth]{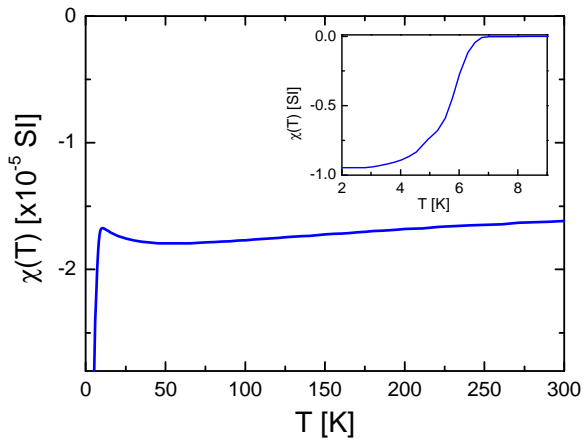}  
\caption{\label{fig:mT_2}(Color online) ZFC dc-susceptibility versus temperature taken in an applied magnetic field of 3 T. The inset shows the $T$-dependent ZFC dc-susceptibility taken at 1 mT.}
\end{figure}
\section{First Critical Field at 2 K}
The first critical field $B_{c1}$ was evaluated via dc magnetization. In Fig.~\ref{fig:bc1} we show the magnetization vs.\ the applied field taken at 2 K. $B_{c1}$ coincides with the field where the magnetization starts to deviate from a linear behavior. We estimate a $B_{c1}$ value of $1.3 \pm 0.2$ mT.
\section{Temperature dependent susceptibility}
In Fig.~\ref{fig:mT_2} the $T$-dependent ZFC dc-susceptibility $\chi(T)$, measured in a 3 T applied field, is shown. Note that $\chi(T)$ is rather small and negative (i.e., diamagnetic) over the whole temperature range. The little hump near the superconducting transition could be the signature of minor paramagnetic impurities. In the inset of Fig.~\ref{fig:mT_2} we show the low-temperature, low-field ZFC dc susceptibility. The saturation to $-1$ of $\chi(T \rightarrow 0)$ proves the bulk character of superconductivity, as it has been found in samples having the same origin (see data presented in Fig.~3(e) of Ref.~\onlinecite{Mizuguchi2012}).

%

\end{document}